\pdfoutput=1

\documentclass[11pt]{article}

\usepackage[]{acl}
\usepackage[most]{tcolorbox}

\usepackage{times}
\usepackage{latexsym}

\usepackage[T1]{fontenc}

\usepackage[utf8]{inputenc}

\usepackage{microtype}

\usepackage{inconsolata}

\usepackage{amsmath}
\usepackage{booktabs}
\usepackage{multirow}
\usepackage{graphicx}
\usepackage{amssymb}
\usepackage{enumitem}
\usepackage{paralist}

\usepackage{pifont}
\newcommand{\cmark}{\ding{51}}
\newcommand{\xmark}{\ding{55}}

\newcommand{\pairdistill}{\textsc{PairDistill}}
\newcommand{\pairdistills}{\textsc{PairDistill }}

\title{\pairdistill: Pairwise Relevance Distillation for Dense Retrieval}

\author{Chao-Wei Huang \quad Yun-Nung Chen \\
        National Taiwan University, Taipei, Taiwan \\
        \texttt{f07922069@csie.ntu.edu.tw} \quad \texttt{y.v.chen@ieee.org}}

\begin{document}
\maketitle
\begin{abstract}

Effective information retrieval (IR) from vast datasets relies on advanced techniques to extract relevant information in response to queries. 
Recent advancements in dense retrieval have showcased remarkable efficacy compared to traditional sparse retrieval methods. To further enhance retrieval performance, knowledge distillation techniques, often leveraging robust cross-encoder rerankers, have been extensively explored.
However, existing approaches primarily distill knowledge from pointwise rerankers, which assign absolute relevance scores to documents, thus facing challenges related to inconsistent comparisons.
This paper introduces Pairwise Relevance Distillation (\pairdistill) to leverage pairwise reranking, offering fine-grained distinctions between similarly relevant documents to enrich the training of dense retrieval models. 
Our experiments demonstrate that \pairdistills outperforms existing methods, achieving new state-of-the-art results across multiple benchmarks. This highlights the potential of \pairdistills in advancing dense retrieval techniques effectively.\footnote{Our source code and trained models are released at \url{https://github.com/MiuLab/PairDistill}}
\end{abstract}

\section{Introduction}
Information retrieval (IR) is the process of retrieving relevant information from vast datasets, such as web pages or documents, based on user queries. Recently, deep learning methods, notably the dense passage retriever (DPR)~\cite{karpukhin-etal-2020-dense}, have attracted attention for their superior performance compared to traditional sparse retrieval techniques like BM25. These methods, often termed dual-encoder models, encode both queries and documents into high-dimensional representations, facilitating efficient similarity computation and retrieval via nearest neighbor search~\cite{douze2024faiss}.

Despite the effectiveness of dense retrievers, their modeling capacity is limited. To enhance retrieval performance, knowledge distillation is commonly employed~\cite{izacard2020distilling}. Typically, knowledge from a robust cross-encoder reranker is distilled to train the dense retriever, achieving state-of-the-art results on retrieval benchmarks~\cite{santhanam-etal-2022-colbertv2}. The efficacy of knowledge distillation largely relies on the performance of the reranker, which serves as the upper bound for the distilled retriever's performance.

\begin{figure}[t]
    \centering
    \includegraphics[width=\linewidth]{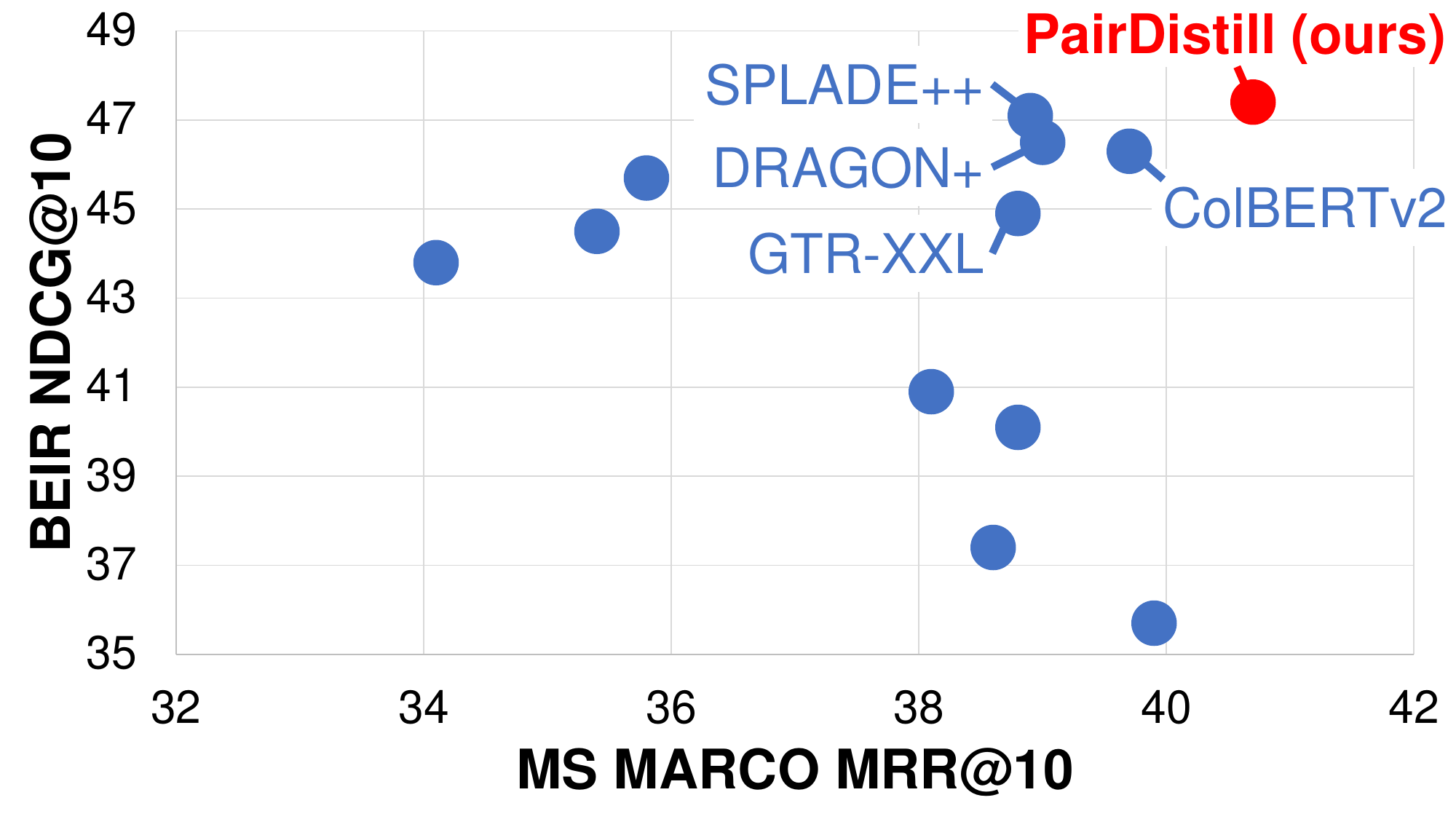}
    \caption{\pairdistill, a model trained with our proposed pairwise relevance distillation, achieves the best performance in both in-domain evaluation (x-axis; MS MARCO dev set) and out-of-domain evaluation (y-axis; average performance over BEIR datasets).}
    \label{fig:scatter}
\end{figure}

However, existing studies primarily utilized \emph{pointwise} rerankers for knowledge distillation, which an absolute relevance score is assigned for each document.
Such scores are not trivial to compare due to inconsistent baselines.
In contrast, \emph{pairwise} reranking, an advanced technique comparing pairs of documents to assess their relative relevance to a query, has demonstrated superior reranking performance~\cite{Pradeep2021TheED}. By emphasizing relative comparison, pairwise rerankers can distinguish more finely between similarly relevant documents, yielding more precise relevance scores conducive to better distillation.

In this paper, we introduce Pairwise Relevance Distillation (\pairdistill), a novel method leveraging the fine-grained training signals provided by pairwise rerankers.
\pairdistills enriches the training of dense retrieval models by distilling knowledge from pairwise comparisons, enabling the model to learn more nuanced distinctions between closely ranked passages.
We conduct extensive experiments and demonstrate that \pairdistills outperforms all baseline of similar size on multiple benchmark, as shown in Figure~\ref{fig:scatter}.
In addition, we show that \pairdistills is effective across difference architectures, i.e., ColBERT\cite{khattab2020colbert} and DPR~\cite{karpukhin-etal-2020-dense}, and in a domain adaptation setting.
Furthermore, we demonstrate the potential of adopting LLM rerankers in \pairdistill.

Our contributions are summarized as follows:
\begin{compactitem}
\item We propose Pairwise Relevance Distillation (\pairdistill), a novel method integrating the advantages of pairwise reranking into dense retrieval model training.
\item Through extensive experiments, we demonstrate that \pairdistills significantly outperforms existing dense retrieval models of similar size.
\item We provide a comprehensive analysis, offering insights into the mechanisms driving the improvements achieved by \pairdistill.
\end{compactitem}

\section{Related Work}

\paragraph{Dense Passage Retrieval}
Dense retrieval has garnered attention for its efficacy in semantic space exploration. A notable technique in this domain is DPR~\cite{karpukhin-etal-2020-dense}, employing both query and passage encoders for efficient retrieval. Various studies have delved into enhancing dense retrieval, including negative example mining techniques like RocketQA~\cite{qu-etal-2021-rocketqa}, and diverse data augmentation methods such as DRAGON~\cite{lin2023train}. ColBERT~\cite{khattab2020colbert,santhanam-etal-2022-colbertv2} introduced the late-interaction mechanism, offering an alternative architecture for dense retrieval.

Another line of research is pre-training strategies for dense retrieval.
Approaches like Contriever~\cite{izacard2021contriever}, coCondenser~\cite{gao-callan-2022-unsupervised}, and COCO-DR~\cite{yu-etal-2022-coco} have proposed contrastive pre-training techniques tailored for retrieval tasks. Concurrently, CoT-MAE~\cite{wu2023contextual} and RetroMAE~\cite{xiao-etal-2022-retromae} have focused on masked auto-encoding for pre-training.

As large language models (LLMs) continue to evolve, their integration into dense retrieval has become increasingly widespread. GTR~\cite{ni-etal-2022-large} utilized LLM encoders, showcasing performance gains with increased model size. Similarly, Promptagator~\cite{dai2023promptagator} and InPars~\cite{bonifacio2022inpars} employed LLMs to generate synthetic query-document pairs, effectively enhancing the training of dense retrievers. Building on a similar concept, \cite{huang-etal-2024-unsupervised} extended the approach to multilingual settings, enabling broader applicability.

Our contribution is orthogonal to these studies as we concentrate on refining training signals for knowledge distillation. This suggests that our approach holds potential for integration with other methods to achieve further improvements.

\paragraph{Knowledge Distillation for Dense Retrieval}
Enhancing the performance of dense retrievers often involves employing knowledge distillation techniques. \citet{izacard2020distilling} pioneered the distillation of knowledge from the reader to the retriever, resulting in improved performance in open-domain question answering.
RankDistill~\cite{reddi2021rankdistil} presented a distillation framework for top-k ranking.
Following this, RocketQAv2~\cite{chakrabarty-etal-2022-rocket} and MarginMSE~\cite{hofstaetter2020_crossarchitecture_kd} proposed knowledge distillation from cross-encoder rerankers to enhance dense retrievers, while CL-DRD~\cite{zeng2022curriculum} introduced curriculum learning for cross-encoder distillation. Further advancements include PROD~\cite{lin2023prod}, which proposed a progressive distillation framework, and ABEL~\cite{jiang2023boot}, introducing an alternating distillation framework with impressive zero-shot performance.
These prior work all performed distillation from pointwise rerankers.
On the other hand, our method introduces pairwise relevance distillation, leveraging finer-grained training signals from pairwise rerankers.

\paragraph{Passage Reranking}
Passage reranking serves as a pivotal second-stage process following initial large-scale retrieval efforts. Various studies have introduced deep reranking models that assess the relevance of query-document pairs by encoding them and predicting relevance scores~\cite{nogueira2019passage}.
For instance, MonoT5 \cite{nogueira-etal-2020-document} introduced a generation-based method for passage reranking by fine-tuning LLMs on MS-MARCO \cite{bajaj2016ms}, distinguishing relevant from irrelevant documents. DuoT5 \cite{Pradeep2021TheED} proposed pairwise reranking, simultaneously comparing two documents to significantly enhance reranking performance. TART \cite{asai2022tart} fine-tunes LLMs via multi-task instruction tuning on diverse retriever datasets.

Another line of research focuses on zero-shot passage reranking with LLMs, which removes the need for retrieval supervision. UPR \cite{sachan-etal-2022-improving} pioneered this approach, proposing to rerank passages by estimating the conditional likelihood of generating the query from the passage using LLMs. \citet{huang2024instupr} enhanced reranking performance further by employing instruction-tuned LLMs. Moreover, \citet{sun2023chatgpt} and \citet{ma2023zero} introduced listwise passage reranking by incorporating prompts with ChatGPT.

Our method combines the superior performance of pairwise reranking with knowledge distillation, which improves retrieval performance significantly and results in state-of-the-art performance on multiple benchmarks.

\begin{figure*}[ht]
    \centering
    \includegraphics[width=\textwidth]{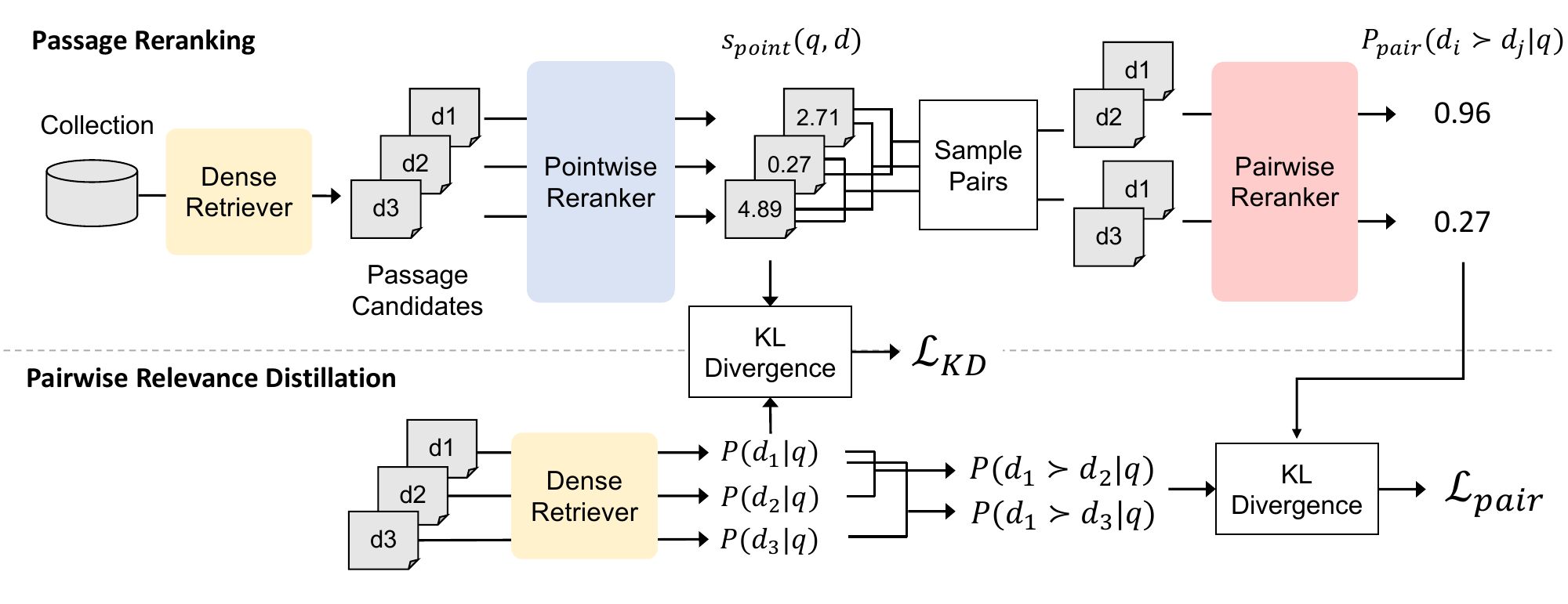}
    \caption{Illustration of our proposed method \pairdistill. \textbf{Top}: The top-k retrieved passages go through pointwise reranking and pairwise reranking to obtain relevance scores. \textbf{Bottom}: Pairwise relevance distillation includes both pointwise distillation loss $\mathcal{L}_{KD}$ and pairwise distillation loss $\mathcal{L}_{pair}$.}
    \label{fig:framework}
\end{figure*}

\section{Background}
In this section, we detail two key tasks: dense retrieval and passage reranking.
Following that, we explore knowledge distillation, a widely adopted technique aimed at bolstering the efficacy of dense retrievers. 
Note that we interchangeably use the terms ``passage'' and ``document'' in this paper.

\subsection{Dense Retrieval}
The goal of dense passage retrieval is to retrieve a subset of relevant passages, denoted as $D^{+}$, from a large collection of passages $\mathcal{D} = \{d_1, \cdots, d_n\}$. 
In order to efficiently retrieve from millions of passages, the most common architecture used for dense retrieval is the dual encoder architecture, where the queries and the passages are encoded by a query encoder and a passage encoder, respectively.
We denote the query representation of a query $q$ as $\mathbf{q}$ and the passage representation of a passage $d$ as $\mathbf{d}$.
This architecture enables offline encoding and indexing of all passages, thus significantly reducing the computation required during retrieval.

The relevance of a query $q$ to a passage $d_i$ is measured using a similarity function:
\begin{equation*}
    \text{s}(q, d_i) = \text{Sim}(\mathbf{q}, \mathbf{d}_i),
\end{equation*}
where a higher similarity score indicates a greater relevance of the passage to the query.
Common choices of the similarity function are dot product, cosine similarity, or the Max-Sum operator introduced in ColBERT~\cite{khattab2020colbert}.

Given a labeled dataset of relevant passage-query pairs $(q, d^{+})$, dense retrievers are typically trained with a contrastive learning objective such as the InfoNCE loss~\cite{oord2018representation}:
\begin{equation*}
    \mathcal{L}_{CL} = -\log \frac{\exp(\text{s}(q, d^{+}))}{\sum_{d \in \mathcal{D}'}\exp(\text{s}(q, d))},
\end{equation*}
where $\mathcal{D}'$ denotes the union of the positive and negative examples.
Optimizing this objective promotes the similarity of the positive pair $s(q, d^{+})$ in contrast to the negative examples.

\subsection{Passage Reranking}
Due to the computational constraints, most dense retrievers utilize lightweight models such as \textit{bert-base}~\cite{devlin-etal-2019-bert} as their backbone model.
Consequently, a subsequent stage of passage reranking aims to refine the initially retrieved passages.
Similar to dense retrieval, the task of passage reranking also aims to assign a relevance score $s_{\text{point}}(q, d_i)$ to each passage $d_i$ given a query $q$.
This reranking scheme is called \textit{pointwise reranking}, where all passages are scored independently.
Given the reduced number of candidate passages at this stage, it becomes feasible to deploy more computationally intensive models.
This allows for the use of cross-encoder architectures and larger models, which are adept at capturing the fine-grained interactions between queries and passages, offering relevance scores that are more accurate.
The candidate passages are then reranked based on their relevance scores $s_{\text{point}}(q, d_i)$.

\subsection{Knowledge Distillation for Dense Retrieval}
Given the success of knowledge distillation of neural models~\cite{hinton2015distilling}, a common approach to enhance the dense retrievers is distilling knowledge from the pointwise rerankers.
Specifically, the relevance of a passage $d_i$ to a query $q$ predicted by a dense retrieval model can be defined as:
\begin{equation*}
    P(d_i \mid q) = \frac{\text{exp} (s(q, d_i))}{\sum_{d \in \mathcal{D}'} \text{exp} (s(q, d))}.
\end{equation*}
Similarly, the relevance predicted by a pointwise reranking model can be defined as:
\begin{equation*}
    P_{\text{point}}(d_i \mid q) = \frac{\text{exp} (s_{\text{point}}(q, d_i) / \tau)}{\sum_{d \in \mathcal{D}'} \text{exp} (s_{\text{point}}(q, d) / \tau)},
\end{equation*}
where $\tau$ is the temperature parameter for controlling the sharpness of the distribution.
Finally, the loss function is the KL divergence between the two distributions:
\begin{equation*}
    \mathcal{L}_{KD} = \frac{1}{|\mathcal{B}|} \sum_{q \in \mathcal{B}} \text{KL}(P_{\text{point}}(d \mid q) \parallel P(d \mid q)),
\end{equation*}
where $|\mathcal{B}|$ denotes the size of the batch.
By optimizing the KL divergence loss, the dense retriever learns to mimic the predictions of the pointwise reranker, thus improving its performance.

\section{Our Method: \pairdistill}
In this section, we introduce our proposed method, pairwise relevance distillation (\pairdistill).
An illustration of the proposed framework is shown in Figure~\ref{fig:framework}.

\subsection{Pairwise Reranking}
\label{sec:pairwise}
While the pointwise rerankers demonstrated superior performance over dense retrievers, reranking all passages independently poses a hard problem in calibrating the relevance score among passages, making the reranking performance of the pointwise rerankers suboptimal.
We conduct preliminary analyses which can be found in Appendix~\ref{app:analyses}.
To mitigate this problem, pairwise reranking techniques can be leveraged.
Pairwise reranking produces better reranking results by comparing two passages simultaneously.

Formally, given a query $q$ and two passages $d_i$ and $d_j$, a pairwise reranker aims to estimate the probability that passage $d_i$ is more relevant to the query than passage $d_j$:
\begin{equation} \label{eq:pairwise}
    s_{\text{pair}}(q, d_i, d_j) = P_{\text{pair}}(d_i \succ d_j \mid q).
\end{equation}
This modeling choice effectively mitigates the calibration problem by only modeling the relative relevance of $d_i$ and $d_j$.
Note that in order to obtain the reranked list, an aggregation method is required which aggregates the relative relevance scores $s_{\text{pair}}(q)$.
However, it is beyond the scope of this paper as our method does not require the final rankings.
In this work, we adopt the following two pairwise reranking methods to estimate the pairwise relevance scores.

\paragraph{Classification-based}
The classification method involves training a binary classifier that predicts whether a given passage $d_i$ is more relevant to a query $q$ than another passage $d_j$.
The classifier takes as input a triplet $(q, d_i, d_j)$ and encodes them together in one sequence, allowing modeling the interaction among the query and two passages.
The output of the classifier will be normalized via a sigmoid function, which can then be interpreted as the probability $P_{\text{pair}}(d_i \succ d_j \mid q)$.
The training objective for this classifier is typically a binary cross-entropy loss, where the model is trained to minimize the difference between the predicted probability and the ground truth relevance ordering of the passages.
This method requires a training dataset consists of triplets and their annotated relative relevance:
\begin{equation*}
    y =
        \begin{cases}
            1 & \text{if $d_i \succ d_j$} \\
            0 & \text{otherwise}
        \end{cases}
\end{equation*}

\paragraph{Instruction-based}
In cases where training data is not available, we adopt instruction-based reranking with LLMs for zero-shot reranking.
We instruct the LLM to select the passage that is more relevant to the query and assign the probability of selecting the index of $d_i$ as the score.
\begin{equation*}
    P_{\text{pair}}(d_i \succ d_j \mid q) = P_{\text{LLM}}(i \mid q, d_i, d_j),
\end{equation*}
where $P_{\text{LLM}}(i \mid q, d_i, d_j)$ is the probability predicted by the LLM of $d_i$ being more relevant to the query $q$ than $d_j$.
The detailed instructions for this method can be found in Appendix~\ref{sec:instructions}.

\subsection{Pairwise Relevance Distillation}
Given the pairwise relevance scores from the pairwise reranker, we can leverage knowledge distillation to further enhance the performance of the dense retriever.
The goal is to make the dense retriever imitate the output distribution of the pairwise reranker, which is defined above in Equation~\ref{eq:pairwise}.
Specifically, we define the pairwise relevance distribution predicted by the dense retriever as:
\begin{equation*}
    P(d_i \succ d_j \mid q) = \\
    \frac{\text{exp} (s(q, d_i))}{\text{exp} (s(q, d_i)) + \text{exp} (s(q, d_j))},
\end{equation*}
which applies the softmax function to the individual relevance scores $s(q, d_i)$ and $s(q, d_j)$.
Consequently, the training objective for pairwise relevance distillation is defined as the KL divergence between the pairwise relevance distributions from the dense retriever and the pairwise reranker:
\begin{equation*}
\resizebox{\linewidth}{!}{%
    $
    \begin{aligned}
        \mathcal{L}_{pair} = & \frac{1}{|\mathcal{B}|} \sum_{q \in \mathcal{B}} \Bigg( \sum_{d_i, d_j \sim \mathcal{D}_{pair}} \\
        & \text{KL}\Big(P_{\text{pair}}(d_i \succ d_j \mid q) \parallel P(d_i \succ d_j \mid q)\Big) \Bigg),
    \end{aligned}
    $
}
\end{equation*}
where $\mathcal{D}_{pair} = \{ (d_i, d_j) \mid d_i, d_j \in \text{ret}_{k}(q), i \neq j,\allowbreak |i - j| < \delta \}$ denotes the set of all possible pairs among $\text{ret}_{k}(q)$, which denotes the top-$k$ documents retrieved given the query $q$.
We introduce a simple heuristic, $|i - j| < \delta$, to constrain the possible pairs, where $\delta$ is a hyperparameter.
The intuition is that documents which are ranked further apart are less likely to provide meaningful training signal, as they are already easily distinguishable by the retriever.

In practice, the process begins by using a retriever to retrieve the top-$k$ documents.
These documents are then reranked by a pointwise reranker to refine the ranking and establish the top-$k$ reranked documents.
Finally, we apply pairwise reranking to the pointwise reranked documents, which allows us to derive pairwise relevance scores for the distillation process.
The full loss function is defined as:
\begin{equation*}
    \mathcal{L} = \mathcal{L}_{CL} + \lambda_{KD} \cdot \mathcal{L}_{KD} + \lambda_{pair} \cdot \mathcal{L}_{pair},
\end{equation*}
where $\lambda_{KD}$ and $\lambda_{pair}$ are hyperparameters representing the weight for the distillation losses.
Our proposed method can also be applied to scenarios where no labeled training data is available.
In such cases, the contrastive loss $\mathcal{L}_{CL}$ is discarded:
\begin{equation*}
    \mathcal{L}_{ZS} = \mathcal{L}_{KD} + \lambda_{pair} \cdot \mathcal{L}_{pair}.
\end{equation*}

\subsection{Iterative Training}
To enhance the performance of the retriever and mitigate the risk of overfitting to a static set of top-k passages, we adopt an iterative training strategy.
In each iteration, the retriever trained in the previous iteration is used to build an index and retrieve the top-k documents.
Subsequently, the top-k documents are reranked with pointwise reranking and pairwise reranking, and the trained retriever is fine-tuned with the full loss $\mathcal{L}$.
The fine-tuned retriever then becomes the retriever for the next iteration.
This iterative training allows for refreshing the retrieved documents in each iteration, avoiding training on the fixed set of documents.
Furthermore, the performance of the retriever can be improved iteratively.

\begin{table*}[ht!]
	\centering
  \resizebox{1\textwidth}{!}{
	  	 \setlength\tabcolsep{3pt}
    \begin{tabular}{l|c|cccccccccccc|c|c}

\toprule

\bf Representation & \bf Sparse & \multicolumn{12}{c|}{\bf Dense}& \multicolumn{2}{c}{\bf Mul-vec}\\
\midrule
Model
&\rotatebox[origin=c]{290}{SPLADE++}&
\rotatebox[origin=c]{290}{GTR-XXL}&
\rotatebox[origin=c]{290}{CL-DRD}&
\rotatebox[origin=c]{290}{RocketQAv2}&
 \rotatebox[origin=c]{290}{CoT-MAE}&
 \rotatebox[origin=c]{290}{RetroMAE}&
\rotatebox[origin=c]{290}{coCondenser}&
\rotatebox[origin=c]{290}{Contriever}&
 \rotatebox[origin=c]{290}{DRAGON+}&
 \rotatebox[origin=c]{290}{ABEL-FT}&
 \rotatebox[origin=c]{290}{COCO-DR}&
\rotatebox[origin=c]{290}{GPL}&
 \rotatebox[origin=c]{290}{PTR}&
\rotatebox[origin=c]{290}{ColBERTv2}&
 \rotatebox[origin=c]{290}{PairDistill (Ours)} \\
\midrule
  Pre-training &\cmark &  \cmark& \xmark & \xmark & \cmark & \cmark& \cmark& \cmark& \cmark& \cmark& \cmark& \xmark& \cmark& \xmark& \xmark \\
  \hline
 Distillation&  \cmark   & \cmark &  \cmark &  \cmark &  \xmark& \xmark& \xmark& \xmark& \cmark& \cmark& \xmark& \cmark& \xmark& \cmark& \cmark \\
 \hline
 Target Corpus &  \xmark  & \xmark &  \xmark &  \xmark &  \xmark& \xmark& \xmark& \xmark& \xmark& \cmark& \cmark& \cmark& \cmark& \xmark& \xmark \\
\midrule
\multicolumn{16}{c}{\textbf{MS MARCO (Supervised)}}\\
\midrule
Dev (RR@10)& 38.9& 38.8& 38.1& 38.8& \underline{39.9}$^{\dagger}$ & 35.4& 38.6& 34.1& 39.0 & - & 35.8& -& -& 39.7& \bf 40.7 \\
Dev (R@1K)& 98.2& \bf 99.0& 97.9& 98.1 & 98.5& 97.5& 98.4& 97.9& \underline{98.6} & - & 97.9& -& -& 98.4& 98.5 \\
DL2019& 74.3& -& 72.5& -& 70.0& 68.8& 71.5& 67.8& 74.4 & - & 74.1& -& -& \underline{74.6}& \bf 75.2 \\
DL2020& 71.8& -& 68.3& -& 67.8& 71.4& 68.1& 66.1& 72.3 & - & 69.7& -& -& \bf 75.2& \underline{75.1} \\

\midrule
\multicolumn{16}{c}{\textbf{BEIR (Zero-shot)}}\\
\midrule
TREC-COVID& 71.1& 50.1& 58.4& 67.5& 56.1& \underline{77.2}& 71.2& 59.6& 75.9 & 76.5 & \bf 78.9& 70.0& 72.7& 73.2& 74.2 \\
NFCorpus& 34.5& 34.2& 31.5& 29.3& 32.1& 30.8& 32.5& 32.8& 33.9 & \underline{35.1} & \bf 35.5& 34.5& 33.4& 33.9& 34.5 \\
FiQA-2018& 35.1& \bf 46.7& 30.8& 30.2& 28.3& 31.6& 27.6& 32.9& 35.6 & 34.3 & 31.7& 34.4& \underline{40.4} & 35.6 & 37.1  \\
ArguAna& 52.1& 54.0& 41.3& 45.1& 27.8& 43.3& 29.9& 44.6& 46.9 & \bf 56.9 & 49.3& \underline{55.7} & 53.8 & 45.8 & 46.8 \\
Tóuche-2020& 24.4& 25.6& 20.3& 24.7& 21.9& 23.7& 19.1& 23.0& 26.3 & 19.5 & 23.8& 25.5& \textbf{26.6}& \underline{26.5} & 26.4 \\
Quora& 81.4& \bf 89.2& 82.6& 74.9& 75.6& 84.7& 85.6& 86.5& \underline{87.5} & 84.5 & 86.7& 83.6& -& 85.1 & 85.3 \\
SCIDOCS& 15.9& 16.1& 14.6& 13.1& 13.2& 15.0& 13.7& 16.5 & 15.9 & \bf 17.4 & 16.0& \underline{16.9}& 16.3& 15.5& 16.2 \\
SciFact& 69.9& 66.2& 62.1& 56.8& 60.1& 65.3& 61.5& 67.7& 67.9 & \bf 72.6 & 70.9& 67.4& 62.3& 69.1 & \underline{71.5} \\
NQ& 54.4& \underline{56.8} & 50.0& 50.5& 48.3& 51.8& 48.7& 49.5& 53.7 & 50.2 & 50.5& 48.3& -& 56.3 & \bf 58.3 \\
HotpotQA& \underline{68.6}& 59.9& 58.9& 53.3& 53.6& 63.5& 56.3& 63.8& 66.2 & 65.7 & 61.6& 58.2& 60.4& 67.4 & \bf 69.3 \\
DBPedia& 44.2& 40.8& 38.1& 35.6& 35.7& 39.0& 36.3& 41.3& 41.7 & 41.4 & 39.1& 38.4& 36.4& \underline{44.6} & \bf 46.0 \\
FEVER& \underline{79.6} & 74.0& 73.4& 67.6& 50.6& 77.4& 49.5& 75.8& 78.1 & 74.1 & 75.1& 75.9& 76.2& 79.0 & \bf 80.4 \\
Climate-FEVER& 22.8& \bf 26.7& 20.4& 18.0& 14.0& 23.2& 14.4& \underline{23.7}& 22.7 & 21.8 & 21.1& 23.5& 21.4& 18.2 & 19.4 \\
CQADupStack& 34.1& \bf 39.9& 32.5& -& 29.7& 34.7& 32.0& 34.5& 35.4 & 36.9 & 37.0& 35.7& -& 36.7 & \underline{38.0} \\
Robust04& 45.8& \bf 50.6& 37.7& -& 30.8& 44.7& 35.4& 47.6 & 47.9 & \underline{50.0} & 44.3& 43.7& -& 46.8 & 48.7 \\
Signal-1M& 29.6& 27.3& 28.2& -& 21.1& 26.5& 28.1& 19.9& 30.1 & 28.0 & 27.1& 27.6& -& \underline{30.7} & \bf 31.2 \\
TREC-NEWS& 39.4& 34.6& 38.0& -& 26.1& 42.8& 33.7& 42.8& \underline{44.4} & \bf 45.4 & 40.3& 42.1& -& 42.0 & 41.9 \\
BioASQ& 50.4& 32.4& 37.4& -& 26.2& 42.1& 25.7& 38.3& 43.3 & 45.4 & 42.9& 44.2& -& \underline{52.2} & \bf 54.8 \\

\hline
\hline
Avg. PTR-11 & \underline{47.1}& 44.9& 40.9& 40.1& 35.7& 44.5& 37.4& 43.8& 46.5 & 46.9 & 45.7& 45.5& 45.5& 46.3& \bf 47.4 \\
Avg. BEIR-13 & \underline{50.3}& 49.3& 44.8& 43.6& 39.8& 48.2& 42.0& 47.5& 50.2 & 50.0 & 49.2& 48.6& -& 50.0 & \bf 51.2 \\
Avg. All-18 & 47.4& 45.8& 42.0& -& 36.2& 45.4& 38.9& 44.5& 47.4 & 47.5 & 46.2& 45.9& -& \underline{47.7} & \bf 48.9 \\

\midrule
\multicolumn{16}{c}{\textbf{LoTTE (Zero-shot)}}\\
\midrule
Search (pooled)& 70.9& -& 65.8& 69.8& 63.4& 66.8& 62.5& 66.1& \underline{73.5} & - & 67.5& -& -& 71.4 & \bf 73.9 \\
Forum (pooled)& 62.3& -& 55.0& 57.7& 51.9& 58.5& 52.1& 58.9& 62.1 & - & 56.8& -& -& \underline{63.2} & \bf 65.5 \\
\bottomrule
\end{tabular}
}

 \caption{Retrieval performance on benchmarks (\%). We report NDCG@10 for MS MARCO and BEIR unless otherwise noted. Recall@5 is reported for LoTTE following previous work. The best result for each dataset is \textbf{bolded} and the second best result is \underline{underlined}. $^{\dagger}$The model was trained on a non-standard MS MARCO corpus which includes the title of the passages.}
	\label{tab:main_results}
\end{table*}

\section{Experiments}
Our proposed method, pairwise relevance distillation, can be applied to both supervised datasets and zero-shot domain adaptation tasks.
In this section, we conduct extensive experiments on passage retrieval tasks to validate and analyze the effectiveness of the proposed method.

\subsection{Datasets}
Following previous work, we use MS MARCO~\cite{bajaj2016ms} as the supervised dataset to perform knowledge distillation.
We evaluate our model on the official dev set of MS MARCO.
Additionally, we perform additional evaluation on TREC 19 and 20~\cite{craswell2020overview,craswell2021overview}.
We also perform zero-shot evaluation on BEIR~\cite{thakur2021beir} and LoTTE~\cite{santhanam-etal-2022-colbertv2}.
Detailed description of the datasets can be found in Appendix~\ref{sec:dataset}.

We report evaluation metrics based on the common practice of each dataset: MRR@10 and Recall@1000 for MS MARCO, NDCG@10 for TREC and BEIR, and Success@5 for LoTTE.

\subsection{Implementation Details}
We adopt the pretrained ColBERTv2~\cite{santhanam-etal-2022-colbertv2} as the initial retriever with the PLAID engine~\cite{santhanam2022plaid} using their official implementation\footnote{\url{https://github.com/stanford-futuredata/ColBERT}}.
Following ColBERTv2, we employ MiniLM\footnote{\url{https://huggingface.co/cross-encoder/ms-marco-MiniLM-L-6-v2}} as the pointwise cross-encoder reranker~\cite{thakur2021beir}, which achieves comparable performance as MonoT5-3B\footnote{\url{https://huggingface.co/castorini/monot5-3b-msmarco}} in our preliminary experiment.
We adopt duoT5-3B\footnote{\url{https://huggingface.co/castorini/duot5-3b-msmarco}}~\cite{Pradeep2021TheED} as our pairwise reranker, which is trained on MS MARCO.
We will discuss the feasibility of using instruction-based reranking with LLMs in Section~\ref{sec:adaptation}.

We control the computational costs for pointwise and pairwise reranking to be the same in our experiments.
For each query, we retrieve top-100 passages from the MS MARCO collection and perform pointwise reranking.
We sample 50 pairs of passages from all possible pairs and obtain pairwise relevance scores through pairwise reranking.
We use all 800K queries for knowledge distillation, while the 500K labeled queries are used for contrastive learning.
$\delta$ is set to 10 in our experiments.
All experiments are conducted with 4 V100 GPUs with 32GB memory each.
Detailed hyperparameters can be found in Appendix~\ref{sec:parameters}.

\begin{table}[ht]
\centering
\begin{tabular}{lccc}
\toprule
 & NQ & TriviaQA & SQuAD \\ 
\midrule
BM25 & 44.6 & 67.6 & 50.6 \\ 
SPLADEv2 & 65.6 & 74.7 & 60.4 \\ 
ColBERTv2 & 68.9 & 76.7 & 65.0 \\ 
\pairdistill & \textbf{71.8} & \textbf{77.4} & \textbf{66.9} \\ 
\bottomrule
\end{tabular}
\caption{Recall@5 performance on open-domain question answering datasets (\%).}
\label{tab:odqa}
\end{table}

\subsection{Main Results}
We compare the performance of our proposed \pairdistills to various baseline models, including state-of-the-art models, e.g., SPLADE++, ColBERTv2, DRAGON+, and ABEL-FT.
The evaluation results on MS MARCO, BEIR, and LoTTE are shown in Table~\ref{tab:main_results}. 
Note that we follow~\citeauthor{lin2023train} and compare with models trained on MS MARCO without title for a fair comparison.

\subsubsection{In-domain Evaluation}
Following previous work~\cite{santhanam-etal-2022-colbertv2,lin2023train,jiang2023boot}, we consider MS MARCO dev set, TREC DL19 and DL20 as in-domain evaluation sets.
As shown in Table~\ref{tab:main_results}, our proposed method \pairdistills achieves 40.7 in terms of MRR@10, which is the best performance on MS MARCO Dev set.
Our model significantly outperforms ColBERTv2 (40.7 v.s. 39.7), which is the initialization of our model.
This result demonstrates that the proposed pairwise relevance distillation effectively improves the performance of dense retrievers.
\pairdistills also achieves the best performance on TREC DL19 and the second best performance on TREC DL20.
Note that coCondenser and CoT-MAE are fine-tuned on the MS MARCO passage corpus that has been augmented with title, which makes their performance not directly comparable to our method.

\subsubsection{Out-of-domain Evaluation}
Next, we evaluate the trained model on out-of-domain evaluation dataset to validate its generalizability.
On the BEIR evaluation datasets~\cite{thakur2021beir}, \pairdistills achieves the best overall performance in three different subsets, demonstrating that our model also excels at out-of-domain generalization.
Considering individual datasets, \pairdistills achieves the best performance among all compared models in 6 out of 18 tasks.
Notably, our method outperforms domain-specific models, e.g., ABEL-FT~\cite{jiang2023boot} and Promptagator~\cite{dai2023promptagator}, which leverage the target domain corpus for specialized domain adaptation.
Additionally, our method consistently outperforms ColBERTv2 in 16 out of 18 datasets, showing that pairwise relevance distillation offers consistent out-of-domain improvement.

On the LoTTE evaluation sets~\cite{santhanam-etal-2022-colbertv2}, \pairdistills achieves state-of-the-art performance in both search and forum subsets, significantly outperforms all compared models.
Notably, DRAGON+~\cite{lin2023train} performs comparably to our model in the search subset, which shows that diverse data augmentation might further improve our model in this scenario.

We also evaluate our model on open-domain question answering datasets, i.e., NaturalQuestions~\cite{kwiatkowski-etal-2019-natural}, TriviaQA~\cite{joshi-etal-2017-triviaqa}, and SQuAD~\cite{rajpurkar-etal-2016-squad}.
We follow ColBERTv2~\cite{santhanam-etal-2022-colbertv2} which reports the performance on the dev set of each dataset in terms of Recall@5.
The results are reported in Table~\ref{tab:odqa}.
\pairdistills consistently outperforms all baseline models on all datasets, demonstrating that our method is suitable for retrieving passages for open-domain question answering as well.

\begin{table}[ht]
\centering
\begin{tabular}{l|c}
\toprule
 & \multicolumn{1}{c}{\textbf{MS MARCO Dev}} \\ 
\midrule
\multicolumn{2}{c}{Distillation Loss} \\
\midrule
PairDistill & \textbf{40.7} \\ 
\quad - $\mathcal{L}_{pair}$  & 39.7 \\
\quad - $\mathcal{L}_{KD}$ & 39.4 \\
\quad - pair sampling heuristic & 40.3 \\
\midrule
\multicolumn{2}{c}{Initialization} \\
\midrule
ColBERTv2 & \textbf{40.7} \\
bert-base-uncased & 40.3 \\
\midrule
\multicolumn{2}{c}{Different Architecture} \\
\midrule
DPR                                    & 34.8                             \\
\quad + $\mathcal{L}_{KD}$             & 36.1                             \\
\quad + $\mathcal{L}_{KD} + \mathcal{L}_{pair}$ & \textbf{36.8}    \\ 
\midrule
\multicolumn{2}{c}{Iterative Training} \\
\midrule
Iteration 1                                    & 40.2                             \\
Iteration 2            & \textbf{40.7}                             \\
Iteration 3            & \textbf{40.7}                             \\
\bottomrule
\end{tabular}

\caption{Results of ablation studies. We report performance on MS MARCO dev set by removing components of our proposed method.}
\label{tab:ablation}
\end{table}

\section{Discussions}

\subsection{Ablation Study}
We conduct ablation studies on MS MARCO dev set to assess the effectiveness of each component in \pairdistill.
Table~\ref{tab:ablation} shows the results of the ablation studies.

In the first experiment, we remove each distillation loss during training.
Note that ColBERTv2 can be seen as an ablation where there’s no $\mathcal{L}_{pair}$.
Removing both $\mathcal{L}_{pair}$ and $\mathcal{L}_{KD}$ results in degraded performance.
Notably, training with only $\mathcal{L}_{pair}$ slightly hurts performance.
Our hypothesis is that since our pairwise distillation objective effectively demotes the score of the lower-ranked passage, we might demote the passage too much during training if we do not refresh the top-k passages.
We also remove the heuristic for pair sampling, where we sample from all possible pairs.
Removing the heuristic shows slight degradation, demonstrating the heuristic contributes to the improvement.

Next, as ColBERTv2 is an already well-trained model, we train our model with different initializations to verify if our method is effective for other pretrained models.
As the results demonstrate, initializing our model with bert-base-uncased achieves 40.3 on MS MARCO dev set.
This result shows that our method is effective regardless of the initialization.

Our proposed method is agnostic to the architecture used for dense retrieval as long as it produces a relevance score for each query-passage pair.
Therefore, we conduct experiments with a different dense retrieval architecture, i.e., DPR~\cite{karpukhin-etal-2020-dense}, to verify if the improvement is consistent across different architectures.
Experimental results shows consistent improvement over vanilla DPR, where using both pointwise and pairwise distillation losses achieves the best performance.
This result demonstrates that our proposed method can improve performance across different dense retrieval architectures.

Finally, we evaluate our trained models from each iteration to verify the effectiveness of the iterative training framework.
The result shows that we can achieve state-of-the-art performance with only 1 iteration, while the second iteration further improves the result.
The improvement converges after 2 iteration.

\begin{table}[t]
\begin{tabular}{lrrr}

\toprule
            & \multicolumn{1}{l}{FiQA} & \multicolumn{1}{l}{BioASQ} & \multicolumn{1}{l}{C-FEVER} \\
\midrule
ColBERTv2   & 35.6                     & 52.2                       & 18.2                        \\
PairDistill & 37.1                     & 54.8                       & 19.4                        \\
\midrule
\multicolumn{4}{c}{Domain Adaptation} \\
\midrule
$\mathcal{L}_{KD}$ only  & 38.2                     & 57.0                         & 21.4                        \\
$\mathcal{L}_{pair}$       & \bf 39.5                     & \bf 59.4                       & \bf 22.6 \\                      
\bottomrule
\end{tabular}
\caption{Performance of zero-shot domain adaptation on FiQA, BioASQ, and Climate-FEVER.}
\label{tab:adaptation}
\end{table}

\subsection{Zero-shot Domain Adaptation}
\label{sec:adaptation}
As discussed in Section~\ref{sec:pairwise}, it is possible to leverage LLMs to perform zero-shot instruction-based reranking.
In this section, we conduct a study where we utilize LLMs for zero-shot domain adaptation.
Specifically, we replace the supervised rerankers with LLMs (flan-t5-xl) for instruction based pointwise and pairwise reranking.

To evaluate the effectiveness of zero-shot domain adaptation with LLMs, we select 3 datasets from BEIR, FiQA, BioASQ, and Climate-FEVER, where training queries are available.
Note that our method only utilize the queries, not the labeled pairs.
We fine-tune ColBERTv2 with $\mathcal{L}_{ZS}$ on each dataset and evaluate the models on the corresponding test set.

Table~\ref{tab:adaptation} shows the results of zero-shot domain adaptation.
Training with $\mathcal{L}_{pair}$ consistently improves performance in the target domain compared to using $\mathcal{L}_{KD}$ only and the baseline models trained on MS MARCO only.
The results demonstrate that performing domain adaptation on queries from the target domain with LLMs is effective.

\section{Conclusion}
In this paper, we introduce Pairwise Relevance Distillation (\pairdistill), a novel distillation method for dense retrieval that leverages the finer-grained training signal provided by the pairwise rerankers.
Through extensive experiments, we demonstrate that \pairdistills achieves state-of-the-art performance in both in-domain and out-of-domain evaluation.
Further analyses show that the proposed method offers consistent improvements across domains and architectures.
We hope this study could provide insights into distillation methods for dense retrieval and prompt more advance distillation techniques.

\section{Limitations}
While the proposed method leverages pairwise relevance for enhancing the training of dense retrievers, it is important to acknowledge certain limitations. One notable concern is the potential requirement for a larger number of training pairs compared to methods utilizing pointwise relevance. This reliance on a larger volume of training pairs may pose challenges in terms of computational resources required for training.

Therefore, future work in this domain should focus on addressing this limitation by exploring strategies to mitigate the need for an extensive number of training pairs while maintaining or even improving the effectiveness of knowledge distillation. This could involve investigating techniques to optimize the selection of training pairs to reduce the computational cost. Addressing the challenge of reducing the required training pairs for knowledge distillation would contribute to the scalability and applicability of the proposed method in real-world retrieval scenarios.

\section*{Acknowledgements}
We thank the reviewers for their insightful comments.
This work was financially supported by the National Science and Technology Council (NSTC) in Taiwan, under Grants 111-2222-E-002-013-MY3 and 112-2223-E002-012-MY5. 
We thank the National Center for High-performance Computing (NCHC) of National Applied Research Laboratories (NARLabs) in Taiwan and HelperAI for providing computational and storage resources.

\bibliography{anthology,custom}

\appendix

\begin{table}[ht]
\centering
\begin{tabular}{lc}
\toprule
\textbf{MSMARCO}   & \textbf{MRR@10} \\
\midrule
ColBERTv2          & 39.7            \\
MiniLM (pointwise) & 40.5            \\
MonoT5 (pointwise) & 40.6            \\
duoT5 (pairwise)   & 41.5            \\
\bottomrule
\end{tabular}
\caption{Reranking performance of different rerankers (\%).}
\label{tab:reranking}
\end{table}

\begin{table}[ht]
\centering
\begin{tabular}{l|c}
\toprule
$\lambda_{pair}$ & \multicolumn{1}{c}{\textbf{MS MARCO Dev}} \\ 
\midrule
1.0 & 40.3 \\
3.0 & 40.7 \\
\bottomrule
\end{tabular}

\caption{Results of varying the value of $\lambda_{pair}$.}
\label{tab:lambda}
\end{table}

\section{Additional Analyses}
\label{app:analyses}

\subsection{Reranking Performance}
In order to better motivate the proposed method, we compare the reranking performance of the pairwise reranker to pointwise rerankers.
Results are shown in Table~\ref{tab:reranking}.
The results demonstrate that pairwise reranking offers greater reranking performance, which makes better distillation targets.

\subsection{Difference between pairwise and pointwise reranking}
In addition to the reranking performance, we conduct another experiment to analyze the difference between pairwise and pointwise rerankers.
In this experiment, we compare the pairwise rank disagreement rate between the rerankers.
We found that the pointwise reranker (MiniLM) disagrees with the more accurate pairwise reranker (duoT5) in 31\% of the pairs sampled via our heuristic.
This result shows that pairwise rerankers provide very different distillation targets for the retrievers.
Combined with the fact that pairwise reranker achieves higher reranking performance, we believe that these experiments demonstrate the necessity of the proposed pairwise relevance distillation.

\subsection{Effect of hyperparameters}
We conduct an experiment where we vary the value of the hyperparameter $\lambda_{pair}$.
The results are shown in Table~\ref{tab:lambda}.
As shown in the results, varying the value of $\lambda_{pair}$ has a slight effect on the final performance.
Setting the value to 3.0 achieves the best performance.

\section{Evaluation Details}

\subsection{Dataset Details}
\label{sec:dataset}
\begin{compactitem}
    \item \textbf{MS MARCO}~\cite{bajaj2016ms}: Following previous work~\cite{santhanam-etal-2022-colbertv2,lin2023train,jiang2023boot}, we use MS MARCO as the supervised dataset, which consists of 502K training queries with 8.8 million passages in the collection. Additionally, there are 306K unlabeled queries that can be used for distillation. The main evaluation is conducted on the official dev set of MS MARCO, which is a standard evaluation set.
    \item \textbf{TREC}~\cite{craswell2020overview,craswell2021overview}: We also perform evaluation on the TREC DL19 and DL20 evaluation sets, which are consider as in-domain datasets as they use the same collection as MS MARCO.
    \item \textbf{BEIR}~\cite{thakur2021beir}: BEIR is a benchmark consisting of 18 retrieval datasets, aiming to assess the out-of-domain retrieval performance of retrievers. We conduct zero-shot evaluation on all 18 datasets.
    \item \textbf{LoTTE}~\cite{santhanam-etal-2022-colbertv2}: LoTTE consists of questions and answers posted on StackExchange with five topics including writing, recreation, science, technology, and lifestyle. A pooled set is also provided where passages and queries from all five topics are aggregated.
\end{compactitem}

\subsection{Baseline Models}
We mostly follow the evaluation procedure from the prior work.
In Table~\ref{tab:main_results}, most results are refered directly from DRAGON~\cite{lin2023train} and ABEL-FT~\cite{jiang2023boot}.
We reran all results of ColBERTv2 to offer a fair comparison to our method.
All evaluation results are computed with the trec\_eval tool from Anserini~\cite{Yang2018AnseriniRR}.

For the open-domain question answering datasets, all baseline results are referred directly from ColBERTv2~\cite{santhanam-etal-2022-colbertv2}.

\subsection{Inference}
During inference, we utilize the PLAID engine~\cite{santhanam2022plaid} for efficient indexing and retrieval.
Following prior work~\cite{santhanam-etal-2022-colbertv2}, we set the maximum length of documents to 300 for BEIR and LoTTE.
The maximum length of queries is set to 300 for Arguana and 64 for Climate-Fever.
We set the compression to 2 bits in the PLAID engine.

\section{Implementation Details}

\subsection{Instruction-based Reranking}
\label{sec:instructions}
For pointwise reranking, we use the following instruction from~\citet{huang2024instupr}:
\begin{tcolorbox}[width=\columnwidth,colback=white]
\small
\begin{verbatim}
Is the document relevant to the query 
(Yes or No)?
Query: {query} 
Document: {document}
\end{verbatim}
\end{tcolorbox}

For pairwise reranking, we use the following instruction:
\begin{tcolorbox}[width=\columnwidth,colback=white]
\small
\begin{verbatim}
Which document is more relevant to the query?
Answer only 'A' or 'B'. 
Query: {query} 
Document: {document}
\end{verbatim}
\end{tcolorbox}

\subsection{Hyperparameters}
\label{sec:parameters}
The hyperparameters used for pairwise relevance distillation training are listed in Table~\ref{tab:parameters}

\begin{table}[ht]
\centering
\begin{tabular}{l|r}
\toprule
\multicolumn{2}{l}{\bf hyperparameters} \\
\midrule
batch size & 32 \\
\# passages per question & 64 \\
max passage length & 180 \\
max query length & 32 \\
max training steps & 100000 \\
learning rate & 1e-5 \\
optimizer & AdamW \\
temperature $\tau$ & 1.0 \\
$\lambda_{KD}$ & 1.0 \\
$\lambda_{pair}$ & 3.0 \\
\bottomrule
\end{tabular}
\caption{Hyperparameters used in the knowledge distillation stage.}
\label{tab:parameters}
\end{table}

\end{document}